\documentclass[doublecol]{epl2}
\usepackage{}
\usepackage{epsfig}
\usepackage{graphicx}
\usepackage{amsfonts,amssymb}
\usepackage{amsmath}
\usepackage{color}
\usepackage {times}
\definecolor{refcolor}{rgb}{1.0,0.0,0.0}
\usepackage{lipsum}
\usepackage[colorlinks,citecolor=blue,linkcolor=blue,anchorcolor=blue,filecolor=blue,urlcolor=blue]{hyperref}

\newcommand{\be}{\begin{equation}}
\newcommand{\ee}{\end{equation}}
\newcommand{\bea}{\begin{eqnarray}}
\newcommand{\eea}{\end{eqnarray}}
\newcommand{\ba}{\begin{array}}
\newcommand{\ea}{\end{array}}

\newcommand{\q}{{\bf q}}
\renewcommand{\k}{{\bf k}}

\begin{document}

\title{Orbital correlations in bilayer nickelates: role of doping and interlayer coupling}

\author{Garima Goyal \and Aastha Jain \and Dheeraj Kumar Singh }

\institute{
\inst{}{Department of Physics and Material Science,
Thapar Institute of Engineering and Technology, Patiala 147004, Punjab, India}\\
}


\date{\today}
\abstract{
We study the nature of orbital correlations present in the  bilayer nickelate within a minimal two-orbital tight-binding model to gain insights into their possible role in stabilizing the less-known weakly-insulating state. The latter has been observed experimentally at ambient pressure. In order to achieve this objective, we examine the static orbital susceptibilities within the random-phase approximation. Our study highlights the sensitivity of orbital correlations to various factors including the interlayer coupling, carrier concentration, bandstructure details such as the orbital contents, the number of bands contributing at the Fermi level etc. We relate this sensitiveness to the modification of the Fermi surfaces as well as their orbital contents dependent on the aforementioned factors.}

\maketitle

\section{Introduction}   
The layered nickelates are the latest addition to the family of unconventional superconductors with cuprates, iron pnictides, and iron chalcogenides~\cite{stewart, bernorz, fernandes} being the prominent members. However, this new class of superconductors are differerent in many ways despite having a layered structure resembling that of cuprates~\cite{karp, werner, gu}. In the case of copper- and iron-based superconductors, the parent compounds exhibit long-range magnetic order and superconductivity appears on doping charge carriers~\cite{taillefer, martinelli, ryan}. On the other hand, parent compounds of infinite-layer~\cite{li, nomura} and bilayer nickelates~\cite{botana, liu} are weakly insulating and paramagnetic metals, respectively. The ground state of undoped bilayer nickelate La$_3$Ni$_2$O$_7$ changes to a weakly insulating state at an elevated pressure $\sim 1$GPa whereas superconductivity appears at a pressure beyond $\sim$ 14GPa~\cite{sun, hou}.

Among various nickelates, La$_3$Ni$_2$O$_7$ exhibits the highest $T_C \sim$ 80K and has therefore attracted considerable attention recently~\cite{sun}. Ni atom being in the octahedral environment, the degeneracy of the five $d$ orbitals is partially lifted, resulting into two sets of orbitals, \textit{i.e.}, $e_g$ and $t_{2g}$. The latter one, lower in energy, is fully occupied~\cite{cao}. Both $e_g$ orbitals contribute at the Fermi level, \textit{i.e.}, $d_{x^2-y^2}$ and $d_{3z^2-r^2}$ are either equally occupied or $d_{x^2-y^2}$ is empty and $d_{3z^2-r^2}$ is fully occupied as the oxidation state of Ni$^{+2.5}$ shows mixed valency~\cite{zhang1}. In a bilayer system, the interlayer coupling between $e_g$ orbitals further leads to $\sigma$-bonding and anti-bonding orbitals~\cite{zhang2}. With the application of pressure, the interlayer coupling can be increased, which will reduce and increase the electronic occupancy in $d_{3z^2-r^2}$ and $d_{x^2-y^2}$ orbitals, respectively, which, in turn, can lead to the appearance of interesting phases~\cite{liu1}.

The electronic bandstructure of the bilayer nickelates is more similar to the bilayer manganites in comparison to the cuprates. In the cuprates, $d_{3z^2-r^2}$ orbital is $\sim$ 1eV lower in energy with respect to $d_{x^2-y^2}$ orbital~\cite{peng}. On the other hand, partial occupancy of $d_{3z^2-r^2}$ orbital together with $d_{x^2-y^2}$ orbital in manganites makes it a crucial factor to understand a very rich-phase diagram as a function of hole doping~\cite{neupane}. In particular, the manganite can exhibit complex spin-charge-orbital ordered states such as the CE-type, which constitutes ($\pi/2$, $\pi/2$)-type orbital order~\cite{kimura, dagotto, rao, dheeraj, dheeraj1}. The latter is believed to induce charge order and support the ferromagnetic zig-zag chain as well at quarter filling~\cite{tokura, avinash}.

La$_3$Ni$_2$O$_7$ exhibits a weakly insulating state at pressure $\sim$ 1GPa, whose origin as well as nature is of intense debate, and a consensus is yet to emerge. Meanwhile, experiments such as resonant inelastic x-ray scattering (RIXS)~\cite{chen}, nuclear-magnetic resonant (NMR), muon-spin relaxation ($\mu^{+}$ SR)~\cite{chen1} study suggest that a spin-density wave is likely to exist below $\sim$ 150K~\cite{khasanov, dan}, whereas the neutron-scattering experiments do not report any magnetic order down to $\sim$10K~\cite{xie}. Furthermore, RIXS measurements indicate spin-density wave-like order with ordering wavevector ($\pi/2$, $\pi/2$), in contrast with widely discussed ($\pi, 0$)-type spin fluctuations responsible for mediating the pairing mechanism~\cite{zhang3}.

Most of the theoretical studies have largely focused on the instability against magnetic order primarily motivated by the speculation of pairing mediated by spin fluctuations~\cite{cao, lechermann, yang}. They use a tight-binding model based on first-principle calculations. The major features of Fermi surfaces include a nearly circular electron pocket around $(0,0)$, and two concentric hole pockets around ($\pi, \pi$)~\cite{gu1, zhang1}.  Several studies suggest that $\sim (\pi, 0)$ nesting vector may lead to spin fluctuations essential for Cooper pair~\cite{zhang3}, while other suggests ($\pi/2, \pi/2$) nesting vector can be important~\cite{wang}.

In this paper, we explore the nature of orbital correlations in the bilayer nickelate. Because if the orbital correlations are strong enough, then they can play an important role in stabilizing the weakly-insulating state in a manner similar to the zig-zag antiferromagnetic state stabilized by orbital order in manganites. On the other hand, strong enough orbital fluctuations can even act as a binding glue of the Cooper pair for the uncoventional superconductivity as noted previously in the case of iron-based superconductors~\cite{kontani}.

\section{Model and Method}
To study orbital correlations in bilayer nickelate, we consider a two-orbital tight-binding model based on the  $d_{x^2 - y^2}$ and $d_{3z^2 - r^2}$ orbitals as proposed earlier~\cite{luo}. The total Hamiltonian in the presence of interactions is given by
\begin{equation}
    \mathcal{H} = \mathcal{H}_K + \mathcal{H}_I,
\end{equation}
where $\mathcal{H}_{K}$ denotes the delocalization energy gain and the second term $\mathcal{H}_{I}$ is the on-site Coulomb interaction energy.

\begin{figure}[]
    \centering
    \includegraphics[scale = 1.0, width = 8.8cm]{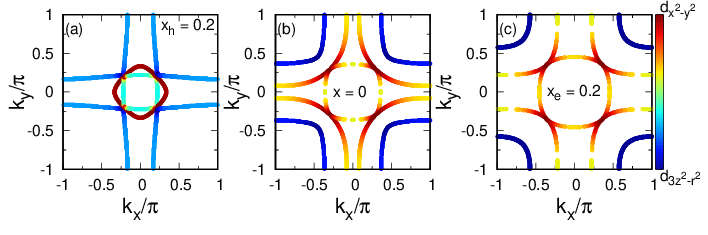}
    \vspace{-10mm}
    \caption{Fermi surfaces in the paramagnetic state with orbital contributions for (a) $x_h = 0.2$, (b) $x_h = 0$, and (c) $x_e = 0.2$.}
    \label{1}
\end{figure}
The kinetic part of the Hamiltonian, because of two layers and two orbital degrees of freedom, takes a 4 $\times$ 4 matrix form
\begin{equation}
    \mathcal{H}_K = \Psi_{{\bf k} \sigma}^{\dagger} H_0 \Psi_{{\bf k} \sigma},
\end{equation}
where
\begin{equation}
    H_0 ({\bf k}) =
    \begin{pmatrix}
        H_{aa} ({\bf k})   &    H_{ab} ({\bf k})   \\
        H_{ba} (\bf k)   &     H_{bb} (\bf k)
    \end{pmatrix}
\end{equation}
and $\Psi_{\sigma} = (d_{a \mu \sigma}, d_{a \nu \sigma}, d_{b \mu \sigma}, d_{b \nu \sigma})$. The subscripts $a$ and $b$ corresponds to top (A) and bottom (B) layers. $\mu$ and $\nu$ refer to the two orbitals $d_{x^2 - y^2}$ and $d_{3z^2 - r^2}$, respectively. $H_{a}$, which is a 2$\times$2 matrix, consists of terms arising due to inter- and intra-orbital hoppings within a single layer as well as on-site energies. It is given by
\begin{equation}
    H_{aa}({\bf k}) = H_{bb}({\bf k}) =
    \begin{pmatrix}
        \varepsilon^{\mu \mu}({\k})    &     \varepsilon^{\mu \nu}({\k}) \\
        \varepsilon^{\nu \mu}({\k})       &   \varepsilon^{\nu \nu} ({\k})
    \end{pmatrix},
\end{equation}
where
\begin{gather}
    \varepsilon^{\mu \mu /\nu \nu}({\k}) = 2t_1^{\mu \mu/\nu \nu} (\cos k_x + \cos k_y) + 4t_2^{\mu \mu/\nu \nu} \cos k_x \cos k_y  \nonumber\\ + \epsilon^{\mu \mu/\nu \nu}, \nonumber \\
    \varepsilon_{\k}^{\mu \nu}({\k}) = 2t_3^{\mu \nu} (\cos k_x - \cos k_y) . \nonumber
\end{gather}
The hopping parameters are $t_1^{\mu \mu} = -0.483$, $t_1^{\nu \nu} = -0.110 $,  $t_2^{\mu \mu } = 0.069 $ $t_2^{\nu \nu} = -0.017$, $t_3^{\mu \nu} = 0.239$~\cite{luo}. The on-site energies are $\epsilon^{\mu \mu } = 0.776$ and $\epsilon^{\nu \nu} = 0.409$.
$H_{ab}$, also a 2$\times$2 matrix, incorporates the interlayer hoppings for different orbitals and it is given by
\begin{equation}
    H_{ab} (\bf k) =
    \begin{pmatrix}
        \varepsilon^{\mu \mu}_{\perp}    &     \varepsilon^{\mu \nu}_{\perp} ({\k}) \\
        \varepsilon^{\nu \mu}_{\perp}({\k})       &   \varepsilon^{\nu \nu}_{\perp}
    \end{pmatrix},
\end{equation}
where
\begin{equation}
         \varepsilon^{\mu \mu}_{\perp} =  t^{\mu \mu}_{\perp}\,{\rm and} \,\,
        \varepsilon^{\mu \nu}_{\perp} ({\k}) = 2t_4^{\mu \nu} (\cos k_x - \cos k_y).
\end{equation}
The hopping parameters are $t^{\mu \mu}_{\perp} =  0.005$, $t^{\nu \nu}_{\perp} = -0.635$, and $t_4^{\mu \nu} = -0.034$~\cite{luo}.

The interaction part of the Hamiltonian is given by
\begin{eqnarray}
\mathcal{H}_{I} &=& U \sum_{{\bf i},l,o} n_{{\bf i} l o  \uparrow} n_{{\bf i} l o \downarrow} + (U' -
\frac{J}{2}) \sum_{{\bf i}, l } n_{{\bf i} l \mu} n_{{\bf i} l \nu} \nonumber \\
&-& 2 J \sum_{{\bf i}, \mu<\nu} {\bf S}_{{\bf i} l \mu} \cdot {\bf S}_{{\bf i} l \nu} + J \sum_{{\bf i},l, \sigma}
d_{{\bf i} l \mu \sigma}^{\dagger}d_{{\bf i} l \mu \bar{\sigma}}^{\dagger}d_{{\bf i} l \nu \bar{\sigma}}
d_{{\bf i} l \nu \sigma},\nonumber\\
\label{int}
\end{eqnarray}
which includes the intra- and inter-orbital Coulomb interaction terms as the first and second terms, respectively. The third term describes the Hund’s coupling and the fourth term represents the pair hopping energy. Rotation-invariant interaction is ensured provided by $U$ = $U^{\prime}$ + $2J$. $l$ is the layer index and $l = A, B$. $o$ is the orbital index and $o = \mu, \nu$.

To study the orbital correlations, we consider the orbital susceptibility defined as follows:
\begin{equation}
\chi^{o}_l(\q,i\Omega_n)= \int^{\beta}_0{d\zeta e^{i \Omega_{n}\zeta}\langle T_\zeta [{\cal O}_{\q l}(\zeta) {\cal O}_{-\q l}(0)]\rangle}.
\end{equation}
$\langle...\rangle$ denotes thermal average, $T_\zeta$ is the imaginary time ordering, and $\Omega_n$ are the bosonic
Matsubara frequencies. ${\cal O}_{{\bf q}l}$ is
obtained as the Fourier transformation of
$\mathcal{O}_{{\bf i}l}$. Since the model considered here has two orbitals, we can define longitudinal and transverse orbital susceptibilities as in the case of spin susceptibility. The relevant longitudinal and transverse orbital operators are given by $\mathcal{O}^{long}_{{\bf i}l} = d^{\dagger}_{{\bf i}l\mu}d_{{\bf i}l\mu} - d^{\dagger}_{{\bf i}l \nu}d_{{\bf i}l\nu}$ and $\mathcal{O}^{trans}_{{\bf i}l} = d^{\dagger}_{{\bf i}l\mu}d_{{\bf i}l\nu} + d^{\dagger}_{{\bf i}l \nu}d_{{\bf i}l\mu}$. Using these definitions, the longitudinal and transverse orbital susceptibility can be obtained via Eq. (8). Thus, it may be  noted that a strong longitudinal orbital correlations may lead to a staggered orbital order involving $d_{x^2-y^2}$ and $d_{3z^2-r^2}$ orbitals. On the other hand, the transverse orbital correlations can lead to a staggered orbital order involving $d^{+}$ and $d^{-}$ orbitals defined by $d^{+} = \frac{1}{\sqrt{2}}(d_{x^2-y^2}+d_{3z^2-r^2})$ and $d^{-} = \frac{1}{\sqrt{2}}(d_{x^2-y^2}-d_{3z^2-r^2})$, or equivalently $d^{+} = d_{z^2-y^2}$ and $d^{-} = d_{x^2-z^2}$~\cite{dheeraj2}.

Because of layer and orbital degrees of freedom, the orbital susceptibility takes a matrix form of size 16 $\times$ 16. Within the random-phase approximation (RPA), the susceptibility $\hat{\chi}^{\rm orb}({\bf q})$ is given by
\begin{eqnarray}
    && \hat{\chi}^{orb}({\bf q}) \!=\!  \hat{\chi}({\bf q})
  [\hat{1}+\hat{U}^{}\hat{\chi}({\bf q})]^{-1},
\label{orb}
  \end{eqnarray}
where $\hat{1}$ is the $16 \times 16$ unit matrix and $\hat{\chi}({\bf q})$ is the bare susceptibility matrix, whose elements can be expressed in terms of Green's function as follows
\bea
&&\chi^{p q, r s}_{\alpha \beta, \gamma \delta}(\q,i\omega_n)  \nonumber\\
&&= \sum_{\k,i\omega^{\prime}_n} G^{p r}_{\alpha \gamma}
(\k+\q,i\omega^{\prime}_n+i\omega_n)G^{s q}_{\delta \beta} (\k,i\omega^{\prime}_n).
\eea
$p$, $q$, $r$, and $s$ are layer indices and take value $1$ and $2$. $\alpha$, $\beta$, $\gamma$, and $\delta$ are the orbital indices and each of them takes two values 1 and 2. It may be noted that all the elements of ${U}^{p q, r s}_{\alpha \beta, \gamma \delta}$ vanish except for some of the elements of ${U}^{1 1, 1 1}_{\alpha \beta, \gamma \delta}$ and ${U}^{2 2, 2 2}_{\alpha \beta, \gamma \delta}$, where ${U}^{1 1, 1 1}_{\alpha \beta, \gamma \delta} = {U}^{2 2, 2 2}_{\alpha \beta, \gamma \delta}$ for all $\alpha$, $\beta$, $\gamma$, and $\delta$. ${U}^{1 1, 1 1}_{\alpha \beta, \gamma \delta} = U, -U^{\prime}+2J, 2U^{\prime}-J,$ and $J$ for $\alpha = \beta = \gamma = \delta$, $\alpha = \gamma \ne \beta = \delta$, $\alpha = \beta \ne \gamma = \delta$, and $\alpha = \delta \ne \beta = \gamma$~\cite{takimoto,singh}.

\begin{figure}[]
    \centering
    \includegraphics[scale = 1.0, width = 8.8cm]{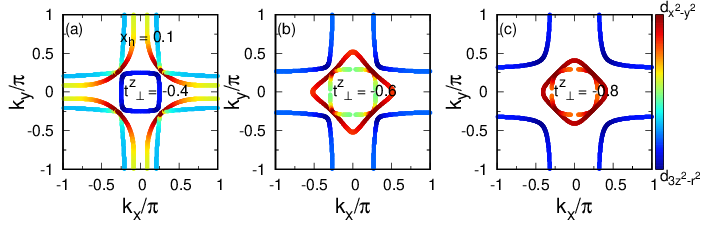}
    \vspace{-8mm}
    \caption{Fermi surfaces in the paramagnetic state with orbital contributions for interlayer hopping of the $d_{3z^2-r^2}$ orbital being $t^z_{\perp}$ = (a) -0.4, (b) -0.6, and (c) -0.8. Here, the hole doping is $x_h$ = 0.1.}
    \label{2}
\end{figure}
\begin{figure}[]
    \centering
    \includegraphics[scale = 1.0, width = 8.8cm]{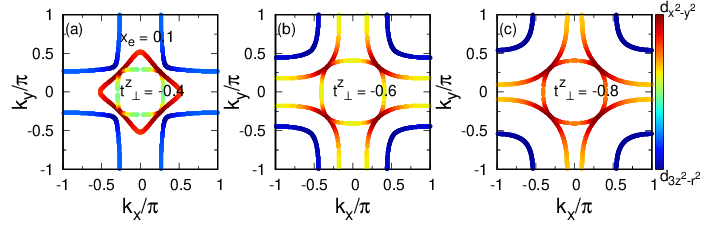}
    \vspace{-8mm}
    \caption{Fermi surfaces in the paramagnetic state with orbital contributions for the interlayer hopping of the $d_{3z^2-r^2}$ orbital being $t^z_{\perp}$ = (a) -0.4, (b) -0.6, and (c) -0.8. Here, the electron doping is $x_e$ = 0.1.}
    \label{3}
\end{figure}

\begin{figure}[]
    \centering
    \includegraphics[scale = 1.0, width = 8.8cm]{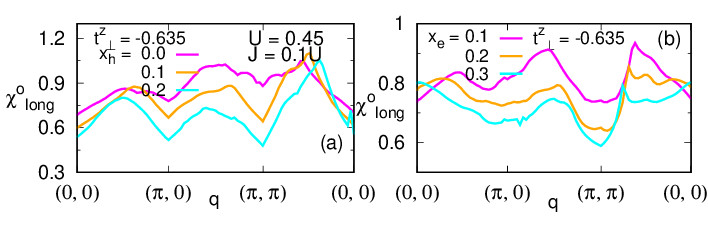}
    \vspace{-8mm}
    \caption{Longitudinal-orbital susceptibility along the high-symmetry directions for various (a) hole and (b) electron dopings. Here $U = 0.44$.}
    \label{4}
\end{figure}

\begin{figure}[]
    \centering
    \includegraphics[scale = 1.0, width = 8.8cm]{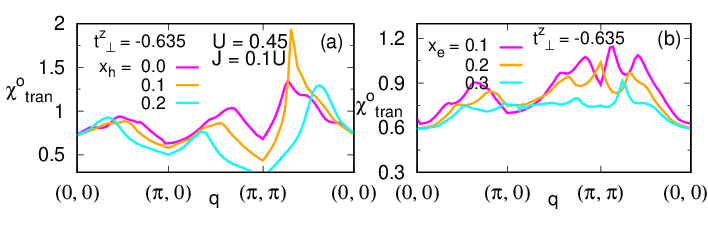}
    \vspace{-8mm}
    \caption{Transverse-orbital susceptibility along the high-symmetry directions for various (a) hole and (b) electron dopings. $U = 0.44$.}
    \label{5}
\end{figure}
\section{Results and discussion}
In the following, we set the unit of energy eV. The Hund's coupling $J$ is set to be $J = 0.1U$ throughout.
The fully occupied $t_{2g}$ orbitals have total 6 electrons at any site, so that the electron density of Ni 3$d$ electrons at a particular site is 7.5, including those from $d_{x^2-y^2}$ and $d_{3z^2-r^2}$ orbitals. In other words, the electronic density of the two $e_g$ orbitals is $n = 1.5$. The chemical potential $\mu = 0$ corresponds to $n = 1.5$, i. e., $x_h = x_e = 0$.

Fig.~\ref{1} (a), (b), and (c) show the Fermi surfaces as a function of dopings. It can be noted that the smaller hole pockets around M$ ((\pi, \pi)) $ is strongly dominated by $d_{3z^2-r^2}$ for $x_e \gtrsim 0$ while the larger one by $d_{x^2-y^2}$. The electron pocket around $\Gamma ((0, 0)) $ has $d_{x^2-y^2}$ orbital predominantly. Along $k_x = k_y$, all the pockets are fully polarized. The size and orbital contents of various pockets depends sensitively on the electronic densities. As the electronic occupancies decrease, the smaller pocket around M becomes larger, its $d_{x^2-y^2}$ content increases and legs are more straightened. On the other hand, there is no significant change in the size of larger pocket except for below $x_h \lesssim 0.1$, where it changes into a smaller pocket, and becomes centered around $\Gamma$, while being mostly dominated by $d_{x^2-y^2}$ orbital. The smaller pocket around $\Gamma$, which is dominated by $d_{x^2-y^2}$ for electron doping, consists of nearly equal contributions from both the $e_g$ orbitals when holes are doped. The major consequence of the changes in the Fermi surfaces, which may affect the nesting, is the increasing distance between the legs of larger and smaller pockets, as the electronic density is increased. This will increase the magnitude of nesting vector.

Another parameter to which the Fermi surfaces and their orbital contents may exhibit sensitiveness is the interlayer coupling $t^{z}_{\perp}$, i.e., the hopping parameter for the $d_{3z^2-r^2}$ orbital in a direction perpendicular to the plane of the layers. This is not surprising because $t^{z}_{\perp}$ is the largest of all the hopping parameters.

Fig.~\ref{2} and \ref{3} show this sensitiveness to $t^{z}_{\perp}$ for $x_h = 0.1$ and $x_e = 0.1$, respectively. Fermi surfaces are plotted for $t^{z}_{\perp} = -0.4$ and $-0.8$ in addition to $t^{z}_{\perp} = -0.6$ a value close to the one originally proposed in the tight-binding model and also considered in the current work. For the hole-doped case (Fig.~\ref{2}), the larger hole pocket can be seen only around $t^{z}_{\perp} \sim -0.4$. For other $t^{z}_{\perp}$, the larger pocket changes to a smaller square-shaped pocket around $\Gamma$ though rotated by $\pi/4$ with respect to the already existing electron pocket. The orbital content of electron pocket around $\Gamma$ is highly sensitive to the interlayer coupling as the nature of orbital dominance get reversed as $t^z_{\perp}$ is changed from $-0.8$ to $-0.4$. At the same time, the legs of the smaller hole pocket around M is further straightened, which can potentially be helpful in improving the nesting properties.

For the electron-doped case, as $t^{z} _{\perp}$ increases, the smaller hole pocket around M becomes more smaller but the change is relatively slow in comparison to the larger hole pocket. On the other hand, the larger hole pocket modifies dramatically to a small pocket surrounding $\Gamma$ when $t^{z}_{\perp}$ becomes smaller than $ \sim -0.4$. There is no significant change in the structure of the electron pocket around $\Gamma$ except for the orbital content. $d_{3z^2-r^2}$ content of the electron pocket continues to increase as $t^{z}_{\perp}$ changes from $\sim -0.8$ to $\sim -0.4$. Just like the hole-doped case, the legs of the pockets around M gets more straightened as $t^z_{\perp}$ is decreased, thus improving the nesting properties.

\begin{figure}[]
    \centering
    \includegraphics[scale = 1.0, width = 8.8cm]{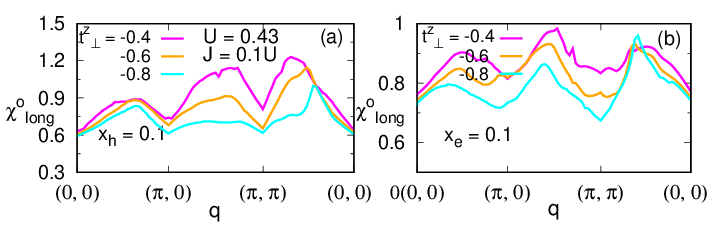}
    \vspace{-8mm}
    \caption{Longitudinal orbital susceptibility along the high-symmetry directions for different interlayer hoppings $t^z_{\perp}$ (a) $x_h = 0.1$ and (b) $x_e = 0.1$.}
    \label{6}
\end{figure}

\begin{figure}[]
    \centering
    \includegraphics[scale = 1.0, width = 8.8cm]{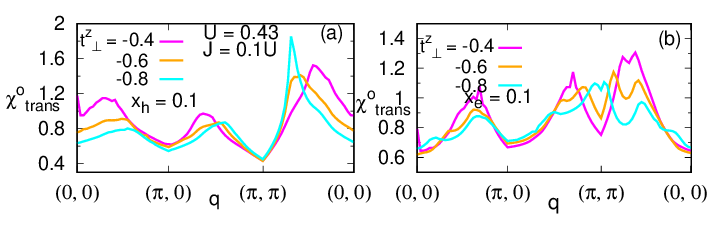}
    \vspace{-8mm}
    \caption{Transverse orbital susceptibility along the high-symmetry directions for different interlayer hoppings $t^z_{\perp}$ (a) $x_h = 0.1$ and (b) $x_e = 0.1$.}
    \label{7}
\end{figure}
Presence of multiple Fermi surfaces may generate several nesting vectors. However, in the bilayer nickelates, it is not difficult to identify the dominant nesting vector. Particularly, the hole pockets around M play very important role as we will see later in this section. The portions of these pockets parallel to $k_x$ and $k_y$ axes give rise to unidirectional as well as bidirectional nesting vectors. There are multiple possibilities even in the unidirectional case because of the interpocket and intrapocket nestings. However, it is the interpocket nesting between the pockets lying on the opposite sides of $k_x = 0$ and $k_y = 0$ lines, which is expected to dominate because there exist two sets of interpocket nesting vectors as compared to one  in the case of intrapocket nesting. Moreover, the bidirectional nesting is expected to prevail over the unidirectional nesting because the same nesting vector will be able to connect all the four sections (around ($\pm \pi, \pm \pi$)) of the hole pockets as compared to only two sections in case of unidirectional nesting.

First, we examine the nature of orbital susceptibility as a function of carrier concentration. Fig.~\ref{4} shows the longitudinal-orbital susceptibility for hole and electron dopings. The largest interaction parameter, i. e., the intraorbital Coulomb interaction is chosen to be $U = 0.44$ as the transverse-orbital susceptibility, to be discussed later, shows diverging behavior beyond it.
For the undoped case, the susceptibility is peaked near the wavevector $\sim (\pi/2, \pi/2)$, $ (\pi, \pi/2)$, and $ (\pi/2, 0)$. However, a relatively stronger peak occurs near $\sim  (\pi/2, \pi/2)$. It may be noted that as the hole doping increases, the magnitude of the wavevectors corresponding to the peak positions also decreases except for $x_h = 0$ when the larger hole pocket disappears. This trend results from a decrease in the distance between the legs of larger and smaller hole pockets around M. An opposite trend is seen when electrons are doped, i. e., the peaks shift towards wavevectors of larger magnitude. Moreover, unlike the hole-doped case, the difference in the peak sizes becomes pronounced especially near $\sim  (\pi, \pi/2)$ and $(\pi/2, \pi/2)$.

Fig.~\ref{5} shows the transverse-orbital susceptibility for different hole and electron dopings. The interaction parameters are the same as in Fig.~\ref{4}. The transverse susceptibility exhibits diverging behavior near $\sim  (\pi/2, \pi/2)$ for $x_h = 0.1$. Note that with a slightly larger $U$, the orbital susceptibility can diverge also for other hole dopings. For the hole doping case, the peaks are sharp and shift towards wavevector of smaller magnitude. The shift towards smaller magnitude of wavevector is similar to that of longitudinal susceptibility and results for similar reasons. Interestingly, $x_h = 0$ curve follows the same trend. On the other hand, when the electrons are doped, the susceptibility becomes increasingly flat, with largest peaks occurring in the vicinity of $\sim  (\pi, \pi)$, which is an overall result of the fact that the peaks are shifting towards the momenta with larger magnitude. Thus, the peaks in the regions X-M and M-$\Gamma$ approach each other as electron concentration continues to increase, where X$\equiv$ ($\pi, 0$). This behavior is in contrast with what we observe for the longitudinal susceptibility, where the peaks in these regions approach each other rather slowly.

Fig.~\ref{6} shows the longitudinal orbital susceptibility as a function of interlayer coupling for (a) hole doping $x_h = 0.1$ and (b) electron doping $x_e = 0.1$. Here, $U \sim 0.43$ as the transverse susceptibility, to be discussed in the next paragraph, shows diverging behavior but the longitudinal susceptibility does not. For $x_h = 0.1$, the susceptibility shows broader peaks near the wavevector $\sim (\pi/2, \pi/2)$, $ (\pi, \pi/2)$, and $ (\pi/2, 0)$ while shifting towards wavevector of smaller magnitude when $t^{z}_{\perp}$ is increased. However, this shift is too small to be noticed near $ \sim (\pi/2, 0)$. Moreover, near the same wavevector, the peak size does not show much change. The largest decline in peak size with increase in $t^{z}$ is noted near $\sim (\pi, \pi/2)$. On the other hand, for $x_e = 0.1$, the peaks shift, upon increasing $t^z_{\perp}$, towards wavevector with smaller magnitude near $\sim (\pi/2, 0)$ and $\sim (\pi, \pi/2)$. No significant shift is noticed in the peak position near $\sim (\pi/2, \pi/2)$. For smallest $t^z_{\perp}$, the peak is highest near $\sim (\pi, \pi/2)$, and for largest $t^z_{\perp}$, the peak is highest near $\sim (\pi/2, \pi/2)$. The peak size decreases with a rise in $t^z_{\perp}$ except near $\sim (\pi/2, \pi/2)$, where it does show significant change.

Fig.~\ref{7} shows the transverse orbital susceptibility for the same set of interaction parameters as in Fig.~\ref{6}. The transverse susceptibility for $x_h = 0.1$ shows diverging behavior for $t^z_{\perp} = -0.8$ near $\sim (\pi/2, \pi/2)$. For all $t^z_{\perp}$s, the peaks near $\sim (\pi/2, \pi/2)$ continue to be the dominant one. All the peaks shift towards wavevector with larger magnitude as $t^z_{\perp}$ is increased. For electron doping $x_e = 0.1$, there is no significant shift in peak position near $\sim (\pi/2, 0)$ whereas the peaks in the regions X-M and M-$\Gamma$ approach each other near M as $t^z_{\perp}$ is increased so that the peak is located near M for $t^z_{\perp} = -0.8$.

\section{Summary and conclusion}
Our findings suggest that the Fermi-surface topology together with the orbital content is very sensitive to the band filling as well as the interlayer coupling, which is well reflected in the shift of the peak positions of the orbital susceptibilities away from or towards $\sim (\pi/2, 0)$, $(\pi, \pi/2)$, and
($\pi/2, \pi/2$). The same is also reflected in the peak size. The shift can be so significant in certain cases that the peaks may come very close to points such as M. For the set of interaction parameters considered here, it may also be noted that the longitudinal susceptibility shows only broader peaks. On the other hand, transverse orbital susceptibility shows relatively sharper peak indicating
that they are on the verge of divergence, and divergence can be achieved by increasing $U$ slightly. Therefore, they can be expected to play an important role in the bilayer nickelates. These orbital fluctuations can be further enhanced through the Jahn-Teller modes associated with the  distortion/vibration of the NiO$_6$ octahedron just like MnO$_6$ octahedron in manganites~\cite{dagotto1}. Such normal modes can also remove the degeneracy of $e_g$ levels.
It may be noted that the strong orbital fluctuations have also the potential to mediate superconductivity, which has been proposed earlier in the case of iron-based superconductors. It was suggested that the sign-preserving $s$-wave superconductivity could be mediated by the orbital fluctuations originating because of the phonon-mediated electron-electron interaction~\cite{kontani1}. In particular, $d_{xz}$, $d_{yz}$, and $d_{xy}$ orbitals were found to be crucial for the ferro-orbital fluctuations. In a similar way, one may expect critically enhanced orbital fluctuations involving partially filled $d_{x^2-y^2}$ and $d_{3z^2-r^2}$ orbitals may mediate superconductivity in this class of materials.

The application of pressure reduces the interatomic distance between the Ni and apical oxygen, thus improving the hopping along the $z$ direction via $d_{3z^2-r^2}$ orbital, an aspect investigated in the current work through the variation of $t^z_{\perp}$. In addition, we have also investigated the role of carrier concentration. A change in carrier concentration can not only affect the Fermi surfaces but it can also potentially modify the nature of ordered state or nature of pairing mechanism. Recent measurements have observed that Pr-doped La$_2$PrNi$_2$O$_7$ show even a larger onset $T_c$ of 82.5 K in comparison to La$_3$Ni$_2$O$_7$~\cite{wang1}. It is not clear yet how the doping of Pr going to change the carrier concentration though  evidence suggests a signature of hole doping~\cite{li}. In future, more experimental works are expected to reveal the consequence of electron or hole doping.

Most of the recent theoretical studies have suggested that the spin-fluctuations may mediate the unconventional superconductivity found in the bilayer nickelates. On the other hand, the current work indicates that the orbital fluctuations, in particular, those transverse in nature, can also be strong in this class of materials. In the presence of both strong spin and transverse orbital fluctuations, it is not unreasonable to expect that the interplay of spin and orbital degrees of freedom can play an important role in stabilizing the weakly insulating state. For instance, such an interplay is known to give rise to interesting but complex phases in other systems such as manganites with the same pair of active 3$d$ orbitals.

To conclude, we have investigated the nature of orbital correlations in a recently proposed two-orbital tight-binding model of bilayer nickelates. Our findings suggest that the transverse orbital correlations with wavevector $\sim (\pi/2, \pi/2$) in this class of superconducting materials are dominant. As a result of the interplay of spin and orbital degrees of freedom, these modes of transverse orbital correlations are expected to play an important role in setting up the less-understood weakly insulating state. The orbital-lattice coupling may further critically enhance this type of orbital fluctuations such that they can even act as a glue for the superconductivity exhibited by this class of materials.

\end{document}